# A Table-top Blast Driven Shock Tube

Michael W. Courtney, Ph.D., U.S. Air Force Academy, 2354 Fairchild Drive, USAF Academy, CO, 80840-6210
Michael.Courtney@usafa.edu

Amy C. Courtney, Ph.D., Force Protection Industries, Inc., 9801 Highway 78, Ladson, SC 29456
amy_courtney@post.harvard.edu

**Abstract:** The prevalence of blast-induced traumatic brain injury in conflicts in Iraq and Afghanistan has motivated laboratory scale experiments on biomedical effects of blast waves and studies of blast wave transmission properties of various materials in hopes of improving armor design to mitigate these injuries. This paper describes the design and performance of a table-top shock tube that is more convenient and widely accessible than traditional compression driven and blast driven shock tubes. The design is simple: it is an explosive driven shock tube employing a rifle primer which explodes when impacted by the firing pin. The firearm barrel acts as the shock tube, and the shock wave emerges from the muzzle. The small size of this shock tube can facilitate localized application of a blast wave to a subject, tissue, or material under test.

## I. Introduction

The prevalence of blast-induced traumatic brain injury in conflicts in Iraq and Afghanistan has motivated laboratory scale experiments on biomedical effects of blast waves and studies of blast wave transmission properties of various materials in hopes of improving armor design to mitigate these injuries.[1][2][3] Compression-driven and blast-driven shock tubes of varying dimensions have been used to apply blast waves to test subjects and materials.

Currently employed shock tube designs suffer from several drawbacks for studying blast-related traumatic brain injury and blast wave transmission, reflection, and absorption in candidate armor materials. Compressed air driven shock tubes exhibit significant shot to shot variations in peak pressure,[4] produce positive pressure durations longer than typically encountered from real threats (antipersonnel mines, hand grenades, improvised explosive devices),[5][6] and often fail to accurately represent the Friedlander waveform of free-field blast waves.[3] Explosive driven shock tubes produce more realistic pressure-time profiles, but their operation requires facilities, liability, and personnel overhead for storing and using high explosive materials. In addition, equipment and personnel need to be isolated from the large mechanical and electromagnetic waves caused by detonation.[1][2] Both air driven and explosive driven shock tubes typically have diameters too large to facilitate isolating exposure to a single anatomical area (head/thorax/abdomen) to isolate injury mechanisms and study wave propagation in animal test subjects and are unwieldy for tissue and cellular level experiments. This paper presents design and characterization of a table-top shock tube that employs firearm primers containing milligram quantities of high explosive and using the firearm barrel as the tube. Unlike existing shock tube designs, this design can apply shock waves with realistic blast wave profiles to small areas of a test subject or candidate armor material.

## II. Method

Rifle primers work by the impact detonation of between 10 and 40 mg of a high-explosive mixture (usually a combination of lead styphnate and lead azide in modern primers), which then ignites the propellant charge.[7] A firearm loaded with a primed cartridge case without any gunpowder or projectile has all the essential elements of an explosive driven shock tube whose blast wave emerges from the muzzle after the primer is detonated by the firing pin. Here, a bolt action rifle chambered in .308 Winchester with a 55.9 cm long barrel is used for the test platform. Tests on large rifle primers employ R-P (Remington) brass cartridge cases with the pockets uniformed and the flash hole deburred. With the method described, this rifle forms a shock tube with a diameter of 7.82 mm. Tests on small rifle primers employ the Lapua Palma cartridge case featuring a small rifle primer pocket. Primers are obtained separately from the cartridge cases and loaded using standard cartridge reloading equipment. Before loading, primer masses are determined with a resolution of 1 mg on an Acculab VIC-123 scale.

Blast pressure measurements employ pressure transducers (PCB 102B and PCB 102B15) placed coaxially with the rifle barrel directly facing the muzzle with no separation between the end of the barrel and pressure transducer. The transducer is connected to a signal conditioning unit (PCB 842C) which produces a calibrated voltage output which is then digitized with a National Instruments PXI-5105 fast analog to digital converter operating at a rate of 1 million samples per second.

| Primer | Peak Pressure (kPa) | SD (kPa) | SD (%) | R | Slope (kPa/mg) |
|---|---|---|---|---|---|
| Fed210M | 2908 | 223 | 7.7% | 0.57 | 73 |
| Fed215M | 3811 | 192 | 5.0% | 0.84 | 64 |
| CCI200 | 2561 | 270 | 10.7% | 0.81 | 124 |
| CCI250 | 3587 | 404 | 11.3% | 0.65 | 130 |
| Rem 7 ½ | 2303 | 186 | 8.1% | 0.66 | 97 |
| Fed205 | 1469 | 103 | 7.1% | 0.03 | N/A |
| CCI450 | 1602 | 104 | 6.5% | 0.59 | 55 |
| Fed205M | 1434 | 103 | 7.2% | 0.69 | 59 |

Table 1: Peak pressure averages, standard deviations from the mean (SD), correlation coefficients (R) with pre-firing total mass, and slopes of peak pressure vs. mass regression line for the table-top shock tube with selected primers. A sample size of ten was used.

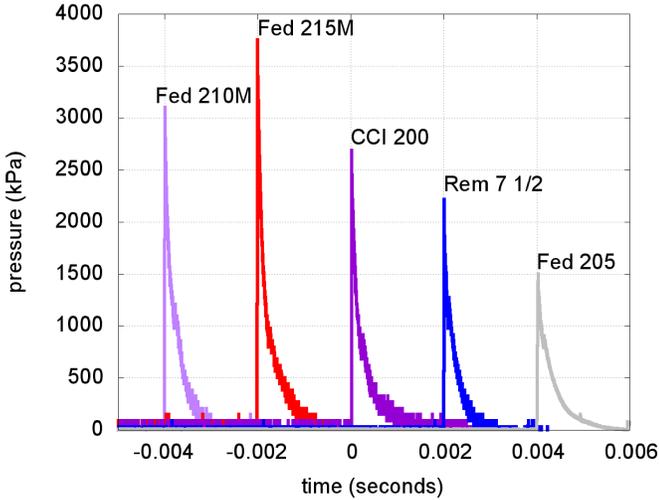

Figure 1: Typical blast pressure waveforms measured for several rifle primer types. The detonation times have been shifted in 0.002 second (2 millisecond) increments to better visualize and compare waveform shapes.

### III. Results

Figure 1 shows blast pressure waveforms for several rifle primer types. Each blast wave is a good approximation to the positive pulse shape of a free-field blast wave demonstrating a rapid rise in pressure followed by a near exponential decay with a time constant of less than 1 ms. Table 1 shows a range of peak pressures are possible from 1.4 MPa to 3.8 MPa by choosing among common commercial rifle primers.

For each primer type (except the CCI 450), peak pressure was well correlated to total primer mass (determined before firing), with correlation coefficients from 0.57 to 0.84 and slopes between 55 and 130 kPa/mg, suggesting a significant reduction in standard deviation is possible by sorting primers by mass to the nearest mg. Testing this idea on a group of ten Fed 210M primers selected to be within 0.5 mg of 355 mg total mass reduced the standard deviation from 7.7% to 4.1%.

Just as peak blast pressure falls off with distance from an explosive source, peak blast pressure also falls off with distance from a shock tube opening. This can be used to apply a smaller blast wave to a sample if desired. The dependence of measured peak blast pressure on the distance from the shock tube opening is shown in Figure 2 for the Federal 210M primer. It is found that the dependence on distance is well represented by a simple model.

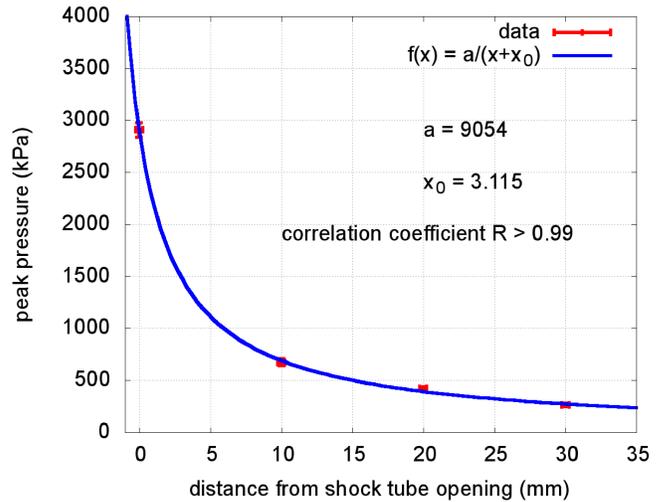

Figure 2: Peak pressure vs. muzzle distance for Fed 210M with best fit model.

### IV. Discussion

The blast wave from the table-top shock tube is of good quality and compares well with free-field blast waves. It should prove useful for experimental designs applying a blast wave over a small area. Using both large and small rifle primers and adjusting the distance from the tube opening from 10 mm to 30 mm, peak magnitudes from 150 kPa to 700 kPa are available. This range spans most of the key region of interest in studying blast-related TBI.[8]

In contrast to free-field blasts, the shock tube described here does not reliably produce a negative pressure rarefaction event following the blast overpressure. Absence of a rarefaction event is common in both larger blast driven [9] and compressed gas driven [4] shock tubes. TBI has been produced in a number of animal models in circumstances where there was no rarefaction following the blast overpressure.[5][10] In addition, the rarefaction is often much smaller than the blast overpressure,[9][3] and is not believed to play a role in proposed TBI mechanisms.

In addition to using different commercial primers and sorting by mass, several variations were tested to adjust the effective area and magnitude of the blast wave. They did not provide undistorted blast profiles, but they are mentioned here to inform others who develop similar systems.

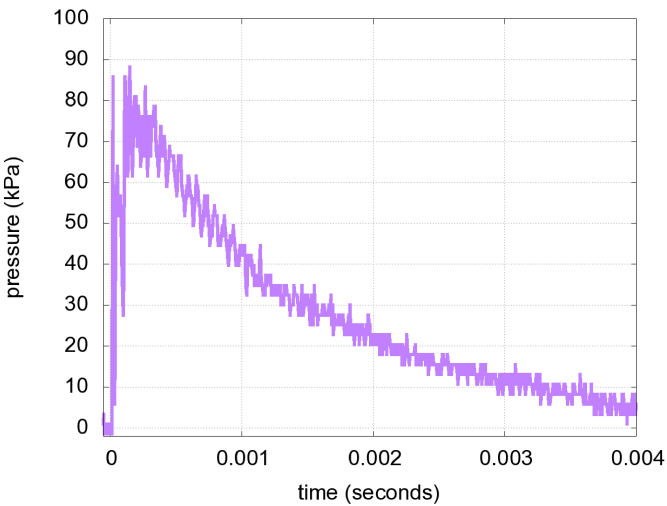

*Figure 3: Pressure vs. time using a Fed 210M primer, and lengthening the tube by inserting the rifle barrel into one end of a steel pipe 1.83 m long and 2.7 cm in diameter.*

Lengthening the tube by connecting the rifle muzzle in one end of larger diameter (2 – 4 cm) steel pipes and measuring the blast wave at the other end reduced the magnitude and increased the area, but distorted the shape and spread the wave out in time over several milliseconds. A typical result is shown in Figure 3. Further experiments would be needed to determine whether a more gradual diameter change and better metal seals (a simple step interface sealed with a vinyl gasket was used here) would preserve a better blast pressure profile.

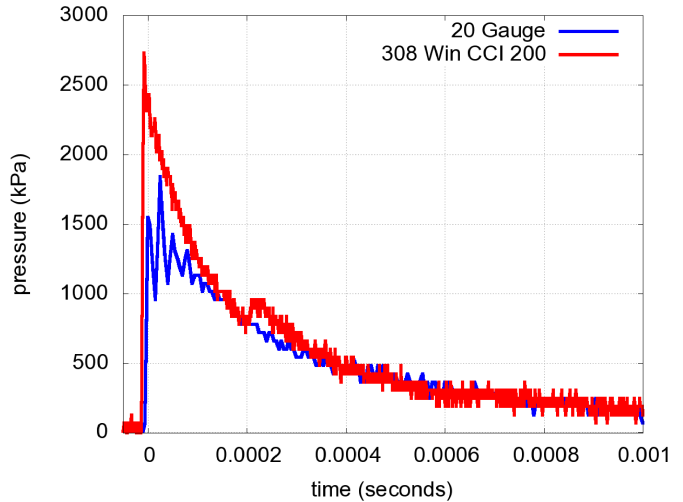

*Figure 4: Pressure vs. time using a 209 shot shell primer measured at the muzzle of a 20 gauge (smaller curve) is jagged and irregular in comparison with the blast wave resulting from a CCI 200 large rifle primer measured at the muzzle of a 308 Winchester rifle (larger curve).*

Blast waves produced with 12 and 20 gauge shotguns were also measured using 209 shotshell primers with no powder, wadding, or shot charge. These produced blast waves with peak pressures of 1800-3000 kPa for tube diameters of 15.63 mm (20 gauge) and 8.53 mm (12 gauge), but the pressure vs. time profiles differed significantly from free-field blast waves. A typical result is shown in Figure 4. Further experiments would be needed to determine at what diameter between the 7.82 mm tube demonstrating a good blast profile and the 15.63 mm tube demonstrating a poor blast profile that the quality of the blast wave begins to break down.

### Acknowledgements:

The authors acknowledge an anonymous RSI reviewer, whose valuable feedback was incorporated into the article.